\documentclass[pdflatex,sn-mathphys-num]{sn-jnl}
\usepackage{graphicx}
\usepackage{multirow}
\usepackage{amsmath,amssymb,amsfonts}
\usepackage{amsthm}
\usepackage{mathrsfs}
\usepackage[title]{appendix}
\usepackage{xcolor}
\usepackage{textcomp}
\usepackage{manyfoot}
\usepackage{booktabs}
\usepackage{algorithm}
\usepackage{algorithmicx}
\usepackage{algpseudocode}
\usepackage{listings}
\begin{document}

\title[Article Title]{The unstable null circular geodesic features the spherically symmetrical asymptotically flat black hole}

\author[1]{\fnm{Yuxuan} \sur{Shi}}\email{shiyx2280771974@gmail.com}

\affil[1]{\orgdiv{Department of Physics}, \orgname{East China University of Science and Technology}, \orgaddress{\city{Shanghai}, \postcode{200237}, \country{China}}}

\abstract{We analyse the massless particles orbiting a spherically symmetric, asymptotically flat black hole with a radius equal to the photon sphere and a circular geodesic. Asymptotic observers record the orbital period of the null circular geodesic as the lowest among all possible paths around the compact object. We proceed with the analytical study of massless particle motion to demonstrate the instability of an orbit with null circularity. Furthermore, our results apply to any asymptotically flat and spherically symmetric spacetime.}


\maketitle

\section{Introduction}
A black hole is thought as a celestial body with an event horizon so enormous that light cannot escape from it, as determined by solving Einstein's field equation\cite{ref1,ref2}. Black holes can now be seen using modern telescopes. The first black hole image, which displayed the black hole shadow of galaxy M87, was taken a few years ago using the Event Horizon Telescope (EHT)\cite{ref3,ref4,ref5,ref6}. Thus, one of the fundamental goals of astrophysics is to comprehend the motion of massless particles in the vicinity of black holes, as this leads to a knowledge of the basic elements of spacetime\cite{ref7,ref8,ref9,ref10,ref11}.

The path of photons near black holes under various impact parameters has been examined in certain research\cite{ref12,ref13,ref14,ref15,ref16}. Additionally, a lower constraint on the effective radius of scalar field hairs outside of black holes is given by circular null geodesics\cite{ref17,ref18,ref19,ref20}. The existence of such a circular null geodesic has been shown by these investigations; for now, however, the general case of a spherically symmetric black hole must be examined. As demonstrated by Hod in Ref.\cite{ref21}, the characteristic of a zero circular geodesic is the shortest possible orbital period observed by the asymptotic observer. This indicates that the equation of a null circular geodesic is equivalent to that of the fastest circular trajectory, or the circular trajectory with the shortest orbital period. In order to analyse these zero-circle orbits and their stability in an asymptotic flat spacetime framework, we employed this as an essential component in our study.

In this paper, we address the null circular geodesic of test particles on the equatorial plane of an asymptotically spherically symmetric black hole. The period of the circular orbit has a minimal value based on Hod's result\cite{ref21}. From the system Lagrangian, we derive and calculate the most minimal effective potential in the radial equation of motion. We further explore the connection between the orbital mass and period of spherically symmetric black holes. Beyond establishing the existence of these geodesics, we also aim to improve our understanding of their behaviour and photon mobility in these kinds of spacetimes.

\section{The null circular orbit and their instability}
The following equation provide the line element of a static, spherically symmetric, asymptotically flat spacetime\cite{ref2}, 
\begin{align}
	\mathrm{d}s^2
	&= g_{\mu\nu}\mathrm{d}x^{\mu}\mathrm{d}x^{\nu}\notag\\
	&= B(r)\mathrm{d}t^2 - A(r)\mathrm{d}r^2 - r^2\left(\mathrm{d}\theta^2 + \sin^2\theta\mathrm{d}\varphi  ^2\right)
\end{align}
where the sole dependence of the metric functions $B(r)$ and $A(r)$ is on the coordinate $r$. The primary noting is that $B(r)A(r)=1$ is not assumed. When $r\to+\infty$, the asymptotic flat spacetime demands that 
\begin{align}
	B(r)\to 1 \quad\text{and}\quad A(r)\to 1
\end{align}
and a horizon radius is the root like,
\begin{align}
	B\left(r_H\right) = A^{-1}\left(r_H\right) = 0
\end{align}
We will presuppose that the particle moves on the equatorial plane, $\theta=\dfrac{\pi}{2}$, as a circle. The Lagrangian is given by\cite{ref22},
\begin{align}
	\mathcal{L} = \dfrac{1}{2}\left[B(r)\dot{t}^2-A(r)\dot{r}^2-r^2\dot{\varphi}^2\right]
\end{align}
where the dots represent the derivative with regard to an affine parameter, which is the appropriate time $\tau$. The generalized momentum are\cite{zuluaga}
\begin{align}
	p_t &= -B(r)\dot t = -E\\
	p_r &= A(r)\dot{r}\\
	p_{\varphi} &= r^2\dot{\varphi} = L
\end{align}
Here, with the existence of two Killing fields $\partial_t$ and $\partial_{\varphi}$, two conserved quantities, named $E$ and $L$, stand for the system's energy and angular momentum, respectively. Eq.(5) and (7) may be solved, allowing us to derive the equations for $\dot{t}$ and $\dot{\varphi}$ as
\begin{align}
	\dot{t} &= \dfrac{E}{B(r)}\\
	\dot{\varphi} &= \dfrac{L}{r^2}
\end{align}
Additionally, the invariant mass for a null particle is zero, meaning that $g_{\mu\nu}\dot{x}^{\mu}\dot{x}^{\nu} = 0$. This suggests
\begin{align}
	\dot{r}^2+V_{\mathrm{eff}}(r) = 0
\end{align}
where the effective potential is given by
\begin{align}
	V_{\mathrm{eff}} = \dfrac{1}{A(r)}\left[\dfrac{L^2}{r^2}-\dfrac{E^2}{B(r)}\right]
\end{align}
The stationary point and zero point of the effective potential (11) are where the equatorial plane's circular orbit is situated, 
\begin{align}
	\dot{r}|_{r=r_c} &= 0\\
	\ddot{r}|_{r=r_c} &= 0
\end{align}
Eq.(12) means $V_{\mathrm{eff}} = 0$, we can find
\begin{align}
	\dfrac{L^2}{r_c^2} = \dfrac{E^2}{B\left(r_c\right)}
\end{align}
The condition (13) requires that $\frac{\mathrm{d}V_{\mathrm{eff}}}{\mathrm{d}r}\big|_{r=r_c} = 0$, which yields
\begin{align}
	\dfrac{\mathrm{d}V_{\mathrm{eff}}}{\mathrm{d}r}\bigg|_{r=r_c}
	&= -\dfrac{2L^2}{r_c^3A\left(r_c\right)}+\dfrac{E^2B'\left(r_c\right)}{B^2\left(r_c\right)A\left(r_c\right)}-\dfrac{L^2A'\left(r_c\right)}{r_c^2A^2\left(r_c\right)}+\dfrac{E^2A'\left(r_c\right)}{B\left(r_c\right)A^2\left(r_c\right)}\notag\\
	&= -\dfrac{2L^2}{r_c^2A\left(r_c\right)}+\dfrac{B'\left(r_c\right)}{B\left(r_c\right)A\left(r_c\right)}\dfrac{L^2}{r_c^2}\notag\\
	&= 0
\end{align}
The second step involves substituting Eq.(14) into Eq.(15) and the above equation simplifies
\begin{align}
	-\dfrac{2}{r_c}+\dfrac{B'\left(r_c\right)}{B\left(r_c\right)} = 0
\end{align}
To ascertain the stability of the inner orbit, one must get the second derivate of the effective potiential (11) at $r=r_c$\cite{ref24}.
\begin{align}
	\dfrac{\mathrm{d}^2V_{\mathrm{eff}}}{\mathrm{d}r^2}
	&= \dfrac{6L^2}{r_c^4A\left(r_c\right)}-\dfrac{2E^2\left[B'\left(r_c\right)\right]^2}{B^3\left(r_c\right)A\left(r_c\right)}+\dfrac{4L^2A'\left(r_c\right)}{r_c^3A^2\left(r_c\right)}\notag\\
	&\quad-\dfrac{2E^2B'\left(r_c\right)A'\left(r_c\right)}{B^2\left(r_c\right)A^2\left(r_c\right)}+\dfrac{2L^2\left[A'\left(r_c\right)\right]^2}{r_c^2A^3\left(r_c\right)}\notag\\
	&\quad-\dfrac{2E^2\left[A'\left(r_c\right)\right]^2}{B\left(r_c\right)A^3\left(r_c\right)}+\dfrac{E^2B''\left(r_c\right)}{B^2\left(r_c\right)A\left(r_c\right)}\notag\\
	&\quad-\dfrac{L^2A''\left(r_c\right)}{r_c^2A^2\left(r_c\right)}+\dfrac{E^2A''\left(r_c\right)}{B\left(r_c\right)A^2\left(r_c\right)}
\end{align}
by substituting Eq.(14) and (16), is becomes
\begin{align}
	\dfrac{\mathrm{d}^2V_{\mathrm{eff}}}{\mathrm{d}r^2}\bigg|_{r=r_c} = \dfrac{E^2}{B\left(r_c\right)A\left(r_c\right)}\left[-\dfrac{2}{r_c^2}+\dfrac{B''\left(r_c\right)}{B\left(r_c\right)}\right]
\end{align}
The period of a circular orbit for a null particle on the equatorial plane is found by using the fact that Hod shown in Ref.\cite{ref21}, 
\begin{align}
	T(r) = \dfrac{2\pi r}{\sqrt{B(r)}}
\end{align}
with $\mathrm{d}s=\mathrm{d}r=\mathrm{d}\theta=0$ and $\Delta\varphi=2\pi$. Eq.(19) makes clear that, in comparison to the classical situation, the orbital period adds a correction term to the denominator, which is the redshift factor. It possesses $T\left(r\to r_H\right)\to+\infty$ and $T\left(r\to+\infty\right)\to+\infty$ under the asymptotically flat conditions, $r\to r_H$ and $r\to+\infty$. Consequently, by Rolle's theorem, there must exist a radius $r_c$ such that $T(r)$ has an extremum, $\frac{\mathrm{d}T}{\mathrm{d}r}\big|_{r=r_c}=0$. Therefore, we can get
\begin{align}
	\dfrac{\mathrm{d}T}{\mathrm{d}r}\bigg|_{r=r_c} = \dfrac{2\pi}{\sqrt{B\left(r_c\right)}}-\dfrac{\pi r_cB'\left(r_c\right)}{B^{\frac{3}{2}}\left(r_c\right)}=0
\end{align}
or
\begin{align}
	\dfrac{2}{\sqrt{B\left(r_c\right)}}\left[1-\dfrac{r_cB'\left(r_c\right)}{2B\left(r_c\right)}\right]=0
\end{align}
Of the radius $r_c$, the period ought to be the smallest. Hence, at least one real root must exist for Eq.(19). This suggests that, in the situation that follows, the period will be lowest at $r=r_c$:
\begin{align}
	\dfrac{\mathrm{d}^2T}{\mathrm{d}r^2}\bigg|_{r=r_c} > 0
\end{align}
From Eq.(19) we have
\begin{align}
	\dfrac{\mathrm{d}^2T}{\mathrm{d}r^2}\bigg|_{r=r_c} = -\dfrac{2\pi B'\left(r_c\right)}{B^{\frac{3}{2}}\left(r_c\right)}+\dfrac{3\pi r_c\left[B'\left(r_c\right)\right]^2}{2B^{\frac{5}{2}}\left(r_c\right)}-\dfrac{\pi r_cB''\left(r_c\right)}{B^{\frac{3}{2}}\left(r_c\right)}
\end{align}
because of $r_c$ satisfying Eq.(19), the latter simplifies to
\begin{align}
	\dfrac{\mathrm{d}^2T}{\mathrm{d}r^2}\bigg|_{r=r_c} = \dfrac{\pi r_c}{\sqrt{B\left(r_c\right)}}\left[\dfrac{2}{r_c^2}-\dfrac{B''\left(r_c\right)}{B\left(r_c\right)}\right]>0
\end{align}
and than we get
\begin{align}
	\dfrac{2}{r_c^2}-\dfrac{B''\left(r_c\right)}{B\left(r_c\right)}>0
\end{align}
Pondering about the part enclosed in brackets to the right of the equal sign in the Eq.(18), we can reveal that the sign of $\frac{\mathrm{d}^2V_{\mathrm{eff}}}{\mathrm{d}r^2}\big|_{r=r_c}$ based on Eq.(25),
\begin{align}
	\dfrac{\mathrm{d}^2V_{\mathrm{eff}}}{\mathrm{d}r^2}\bigg|_{r=r_c}<0
\end{align}
It demonstrates that for any spherically symmetric asymptotically flat black holes, the null circular geodesic is unstable. An example would be a Schwarzschild black hole, $B(r)=A^{-1}(r)=1-\frac{2M}{r}$. The null circular orbit is located at $r_c=3M$\cite{ref25}, and we can obtain the formula below by applying Eq.(18)
\begin{align}
	\dfrac{\mathrm{d}^2V_{\mathrm{eff}}}{\mathrm{d}r^2}\bigg|_{r=r_c} = -\dfrac{2E^2}{3M^2}<0
\end{align}
The outcome of this work shows that it is unstable. Chandrasekhar had also demonstrated the same deduction using a different technique\cite{ref1}. As an additional instance, Mazharimousavi pointed out that the null circular geodesic is unstable for the spherically symmetric, flat, hairy black hole\cite{ref24}. It is not overly dramatic to state that the circular orbit's period must appear as a minimun at $r=r_c$ with meeting Eq.(22). Furthermore, as an outcome of Eq.(22), the $\frac{\mathrm{d}^2V_{\mathrm{eff}}}{\mathrm{d}r^2}\big|_{r=r_c}$ becomes negative, which in turn causes the circle geodesics with a radius of $r_c$ to become unstable. Finally, it is critical to emphasise that $\frac{\mathrm{d}^2V_{\mathrm{eff}}}{\mathrm{d}r^2}\big|_{r=r_c}>0$ and Eq.(22) cannot be met simultaneously. This ultimately results to an unstable orbit around the spherically symmetrical asymptotically flat black holes, in which massless particles travel at $r=r_c$.

\section{Conclusion}

The stability of a circular orbit in the equatorial plane of such a black hole is fundamentally broken, as we demonstrate by the fact that the second derivative of the effective potential is negative. It is noteworthy to point out that this particular type of black hole with an unstable circular orbit also meets the conditions listed below: the circular orbit with a local minimum period of Eq.(22) from Eq.(20) leads to Eq.(27). We also extend the applicability of this conclusion to any spherically symmetric gravitational source that satisfies the necessary prerequisites. The results reported in \cite{ref24}, research the unstability of the geodesic outside a spherically symmetricasymptotically flat hairy black hole, serves only as a particular case. Moreover, we observe that this instability is not limited on the positions $r=r_H$ and $r\to+\infty$. More ways than one, the divergence of the period is made clear in these two occasions: $T\left(r= r_H\right)\to +\infty$ and $T\left(r\to+\infty\right)\to +\infty$. This further underscores the robustness of our interpretation.

\backmatter
\bmhead{Acknowledgements}

The author thanks Hongbo Cheng for helpful discussion.

\end{document}